\documentclass[twocolumn]{aastex63}
\usepackage{lineno}
\shorttitle{Flaring-associated Complex Dynamics in Two M-dwarfs}
\shortauthors{Wang et al.}

\begin{document}

\title{Flaring-associated Complex Dynamics in Two M-dwarfs Revealed by Fast, Time-resolved Spectroscopy}

\correspondingauthor{J. Wang}
\email{wj@nao.cas.cn}

\author{J. Wang}
\affiliation{Guangxi Key Laboratory for Relativistic Astrophysics, School of Physical Science and Technology, Guangxi University, Nanning 530004,
People's Republic of China}
\affiliation{Key Laboratory of Space Astronomy and Technology, National Astronomical Observatories, Chinese Academy of Sciences, Beijing 100101,
People's Republic of China}

\author{H. L. Li}
\affiliation{Key Laboratory of Space Astronomy and Technology, National Astronomical Observatories, Chinese Academy of Sciences, Beijing 100101,
People's Republic of China}

\author{L. P. Xin}
\affiliation{Key Laboratory of Space Astronomy and Technology, National Astronomical Observatories, Chinese Academy of Sciences, Beijing 100101,
People's Republic of China}

\author{G. W. Li}
\affiliation{Key Laboratory of Space Astronomy and Technology, National Astronomical Observatories, Chinese Academy of Sciences, Beijing 100101,
People's Republic of China}

\author{J. Y. Bai}
\affiliation{Key Laboratory of Space Astronomy and Technology, National Astronomical Observatories, Chinese Academy of Sciences, Beijing 100101,
People's Republic of China}

\author{C. Gao}
\affiliation{Guangxi Key Laboratory for Relativistic Astrophysics, School of Physical Science and Technology, Guangxi University, Nanning 530004,
People's Republic of China}
\affiliation{Key Laboratory of Space Astronomy and Technology, National Astronomical Observatories, Chinese Academy of Sciences, Beijing 100101,
People's Republic of China}
\affiliation{School of Astronomy and Space Science, University of Chinese Academy of Sciences, Beijing, People's Republic of China}

\author{B. Ren}
\affiliation{Guangxi Key Laboratory for Relativistic Astrophysics, School of Physical Science and Technology, Guangxi University, Nanning 530004,
People's Republic of China}
\affiliation{Key Laboratory of Space Astronomy and Technology, National Astronomical Observatories, Chinese Academy of Sciences, Beijing 100101,
People's Republic of China}
\affiliation{School of Astronomy and Space Science, University of Chinese Academy of Sciences, Beijing, People's Republic of China}

\author{D. Song}
\affiliation{Guangxi Key Laboratory for Relativistic Astrophysics, School of Physical Science and Technology, Guangxi University, Nanning 530004,
People's Republic of China}
\affiliation{Key Laboratory of Space Astronomy and Technology, National Astronomical Observatories, Chinese Academy of Sciences, Beijing 100101,
People's Republic of China}
\affiliation{School of Astronomy and Space Science, University of Chinese Academy of Sciences, Beijing, People's Republic of China}

\author{J. S. Deng}
\affiliation{Key Laboratory of Space Astronomy and Technology, National Astronomical Observatories, Chinese Academy of Sciences, Beijing 100101,
People's Republic of China}
\affiliation{School of Astronomy and Space Science, University of Chinese Academy of Sciences, Beijing, People's Republic of China}

\author{X. H. Han}
\affiliation{Key Laboratory of Space Astronomy and Technology, National Astronomical Observatories, Chinese Academy of Sciences, Beijing 100101,
People's Republic of China}

\author{Z. G. Dai}
\affiliation{Department of Astronomy, University of Science and Technology of China, Hefei 230026, People's Republic of China}
\affiliation{School of Astronomy and Space Science, Nanjing University, Nanjing, 210023, People's Republic of China}

\author{E. W. Liang}
\affiliation{Guangxi Key Laboratory for Relativistic Astrophysics, School of Physical Science and Technology, Guangxi University, Nanning 530004,
People's Republic of China}

\author{X. Y. Wang}
\affiliation{School of Astronomy and Space Science, Nanjing University, Nanjing, 210023, People's Republic of China}

\author{J. Y. Wei}
\affiliation{Key Laboratory of Space Astronomy and Technology, National Astronomical Observatories, Chinese Academy of Sciences, Beijing 100101,
People's Republic of China}
\affiliation{School of Astronomy and Space Science, University of Chinese Academy of Sciences, Beijing, People's Republic of China}






\begin{abstract}

Habitability of an exoplanet is believed to be profoundly affected by activities of the host stars,
although the related coronal mass ejections (CMEs) are still rarely detected in 
solar-like and late-type stars. We here report an observational study on flares of two M-dwarfs 
triggered by the high-cadence survey performed by the Ground Wide-angle Camera system. In both events,
the fast, time-resolved spectroscopy enables us to identify symmetric broad H$\alpha$ emission 
with not only a nearly zero bulk velocity, but also a large projected maximum velocity as high as $\sim700-800\ \mathrm{km\ s^{-1}}$. This broadening could be 
resulted from either Stark (pressure) effect or a flaring-associated CME at stellar limb.
In the context of the CME scenario, the CME mass is estimated to be 
$\sim4\times10^{18}$ g and $2\times10^{19}$ g. In addition, our spectral analysis reveals a temporal 
variation of the line center of the narrow H$\alpha$ emission in both events. The variation amplitudes are 
at tens of $\mathrm{km\ s^{-1}}$, which could be ascribed to the chromospheric evaporation in one 
event, and to a binary scenario in the other one.
With the total flaring 
energy determined from our photometric monitor, we show a reinforced trend in which larger the flaring energy, higher the CME mass is.

\end{abstract}



\section{Introduction} \label{sec:intro}

It is known for a long time that solar-like and late-type main sequence stars show highly energetic flares.
The flares with total energies of $10^{33-39}$ erg can be detected at multiple wavelengths from radio to 
X-ray (e.g., Pettersen 1989; Schmitt 1994; Osten et al. 2004, 2005; Huenemoerder et al. 2010; Maehara et al. 2012;
Kowalski et al. 2013; Balona 2015; Davenport et al. 2016; Notsu et al. 2016; Van Doorsselaere et al. 2017; Chang et al. 2018;
Paudel et al. 2018; Schmidt et al. 2019; Xin et al. 2021). Given the comprehensive studies on Sun, an analogy with solar 
flares leads to a common knowledge that these stellar flares can be ascribed to stellar magnetic activity, such as magnetic reconnection (e.g., Noyes et al. 1984; Wright et al. 2011; Shulyak et al. 2017). 

Complicated dynamic responses of the chromospheric plasma heated by the energy released in the reconnection is therefore
expected for solar-like and late-type main sequence stars.
On the one hand,  the erupted magnetic field lines can trigger a large scale expulsion of the confined plasma 
into interplanetary space, i.e., a coronal mass ejection (CME) (e.g., Kahler 1992; Tsuneta 1996; Kliem et al. 2000; Karlicky \& Barta 2007; Shibata \& Magara 2011; Li et al. 2016; Jiang et al. 2021),
if the field eruption is strong enough and the overlying fields are not too constraining (see the review by Forbes et al. 2006). 
Stellar CMEs are believed to be essential to the habitability of an exoplanet, which is especially important for M-dwarfs 
since the distance of a habitability zone to the host star is only 0.1AU (Shields et al. 2016).
Simulations, in fact,
suggest that frequent stellar activities can either tear off most of the atmosphere of an exoplanet in long timescale (e.g., Cerenkov et al. 2017; Airapetian et al. 2017; Garcia-Sage et al. 2017) or chemically generate greenhouse gas and HCN in short timescale 
(e.g., Airapetian et al. 2016; Barnes et al. 2016; Tian et al. 2011).
On the other hand, the overpressured chromospheric plasma heated by 
the accelerated electrons
can expand either upward or downward (i.e., chromospheric condensation) with a velocity of $10^{1-2}\ \mathrm{km\ s^{-1}}$ 
in the chromospheric evaporation scenario (e.g.,  Fisher et al. 1985; Milligan et al. 2006a,b; Canfield et al. 1990; Gunn et al. 1994; 
Berdyugina et al. 1999; Antolin et al. 2012; Lacatus et al. 2012; Fuhrmeister et al. 2018; Vida et al. 2019; Li et al. 2019).  

Although both CME and chromospheric evaporation have been observed and studied comprehensively in Sun, 
their detection on solar-like and late-type main sequence stars is still a hard task due to the insufficient spatial resolution 
of contemporary instruments. We refer the readers to Moschou et al. (2019) and Wang et al. (2021) for a brief summary for the  detection of stellar CMEs.
By comparing the solar integrated observations, Namekata et al. (2021) recently reported a probable detection of an eruptive filament from a 
superflare on young solar-type star, EK\, Dra. Basing upon the similar method, Veronig et al. (2021) reported a detection of 21 CME candidates in 
13 late-type stars through X-ray and EUV dimming.  Although Gunn et al. (1994) claimed a detection of chromospheric evaporation with 
maximum velocity of $\sim600\ \mathrm{km\ s^{-1}}$,  the evaporation explanation was argued against recently by 
Koller et al. (2021). A recent case study based on Balmer line asymmetry can be found in Wu et al. (2022). 

In this paper, by following Wang et al. (2021), we report photometric and time-resolved spectroscopic follow-ups of 
flares of two  M-dwarfs triggered by the Ground-based Wide Angle Cameras (GWAC) system.  
The temporal evolution of H$\alpha$ emission line suggests complex dynamics of the heated plasma in both events, that is not only a 
flare-associated CME, but also a possible chromospheric evaporation. 
The remainder of this paper is organized as follows. Section 2
describes the discovery of the two flares. The photometric and
spectroscopic follow-ups, along with the corresponding data reductions, are outlined in Section 3.
Section 4 presents the light-curve and spectral analyses. The
results and discussion are shown in Section 5.

\section{Detection of Flares by GWAC} \label{sec:style}

The GWAC system, including a set of cameras (each with a diameter of 18 cm and a field of view of
150 $\mathrm{deg^2}$) and a set of follow-up telescopes, is one of the ground facilities of
the Space-based multi-band astronomical Variable Objects Monitor (SVOM) mission\footnote{SVOM is a China–France satellite mission dedicated to the detection and
study of gamma-ray bursts (GRBs). Please see Atteia et al. (2022) and the white paper given by Wei et al. (2016) for details.}. 
We refer the readers to Han et al. (2021) for a recent and more detailed description of the GWAC system.

Table 1 tabulates the log of the two flares, i.e., GWAC\,211229A and GWAC\,220106A,  discovered by the GWAC system.  Basically speaking, the two transients without any apparent motion among several 
consecutive images show typical point-spread function (PSF) of nearby bright objects. In fact, there are no known minor planets or
comets\footnote{https://minorplanetcenter.net/cgi-bin/mpcheck.cgi?}  brighter than $V$ = 20.0 mag within a radius of 15\arcmin, and 
no known variable stars or CVs can be found in SIMBAD around the transient positions within 1\arcmin.
In both case, the typical localization error determined from the GWAC images is about 2\arcsec.

For each of the transients, an off-line pipeline involving standard, differential aperture photometry was performed at the location of the transient
and for several nearby bright reference stars using the IRAF\footnote{IRAF is distributed by the National Optical Astronomical Observatories,
which are operated by the Association of Universities for Research in
Astronomy, Inc., under cooperative agreement with the National Science
Foundation.}
APPHOT package, including the corrections of bias, dark, and flat-field. 
A calibration against the SDSS catalog through Lupton (2005) transformations\footnote{http://classic.sdss.org/dr6/algorithms/sdssUBVRITransform.html\#Lupton2005}
is then adopted to obtain the actual brightness of each transient.

\begin{table*}[h!]
\renewcommand{\thetable}{\arabic{table}}
\centering
\caption{Two Optical Transients Discovered by the GWAC System}
\label{tab:decimal}
\begin{tabular}{lcc}
\tablewidth{0pt}
\hline
\hline
Property & GWAC\,211229A  & GWAC\,220106A\\
     (1) & (2) & (3) \\
     \hline
     Trigger Time (UTC)& 11:48:49 & 11:19:28 \\
      R.A. (J2000)           & 23:54:14  & 00:01:32 \\
      DEC (J2000)           & +43:47:23 & +38:41:53\\
      \hline
      \multicolumn{3}{c}{Flaring}\\
      \hline 
      Discovery/Quiescent $R$-band Mag (mag)  & 15.08/16.47  & 15.11/18.64 \\
      Peak $R$-band Mag (mag) &  $11.67\pm0.01$ &  $13.97\pm0.04$\\
      Flaring energy in $R$-band $E_R$ ($\mathrm{erg}$)& $(5.2-5.4)\times10^{33}$ & $(1.2-1.6)\times10^{34}$ \\
      Equivalent Duration, ED (hr) & $6.55-6.83$ & $11.97-16.72$ \\
      \hline
      \multicolumn{3}{c}{Host stars}\\
      \hline
      Quiescent Counterpart & 2MASS\,J23541459+4347232 & 2MASS\,J00013265+3841525\\
      Gaia DR2 ID    &  1922919500519190912 & 2880981530065870720 \\
      $G$-band (mag) &  $15.640\pm0.003$ &  $18.231\pm0.002$\\
      GBp–GRp (mag) &  $3.41\pm0.01$ & $3.01\pm0.07$ \\ 
      Quiescent flux in $R$-band $f_0$ ($\mathrm{erg\ s^{-1}\ cm^{-2}}$) & $4.5\times10^{-16}$ & $6.1\times10^{-17}$ \\
      Distance (pc)     &   $50.6\pm0.2$ & $151.9\pm7.9$\\
      $M_G$ (mag)     &  $12.12\pm0.01$ & $12.32\pm0.11$\\
      $T_{\mathrm{eff}}$ (K)  & $ 2997\pm157$ & $3181\pm160$\\
      $R_\star$  ($R_\odot$) &   $0.220\pm0.007$  & $0.180\pm0.011$ \\
      $M_\star$ ($M_\odot$) &   $0.19\pm0.02$ & $0.15\pm0.02$  \\
      $\mathrm{log} g$ & $5.03\pm0.02$ & 5.1025  \\
      Reference &  \multicolumn{2}{c}{Stassun et al. (2019); Paegert et al. (2021)}  \\
      
\hline

\hline
\end{tabular}
\end{table*}

\section{Follow-up Observations}
\subsection{Photometric Follow-ups and Data Reduction}

Follow-ups in photometry were carried out immediately
by the GWAC-F60A telescope in the standard Johnson–Cousins $R$-band.  
The dedicated real-time automatic transient
validation system (RAVS; Xu et al. 2020) enables us to identify
the transient in minutes and to carry out monitoring with
adaptive sampling that is optimized based upon the
brightness and the evolution trend of each individual target.

With a better localization ($<1\arcsec$) resulted from the follow-up
observations, the quiescent counterparts (or host stars) of the two transients can be identified exactly. 
The properties of the quiescent counterparts quoted from literature are tabulated in Table 1 as well.
With their $G$-band absolute magnitudes and $G_{\mathrm{BP}}-G_{\mathrm{RP}}$ colors, the two host stars
are marked on the color–magnitude diagram (CMD) in Figure 1. Although both stars can be classified as M-dwarfs, 
the host star of GWAC\,211229A is located at the upper boundary of the main sequence, which suggests that the host star of GWAC\,211229A is 
either an unresolved binary or a young main-sequence star (e.g., Gaia Collaboration et al. 2018a,b).

\begin{figure}[ht!]
\plotone{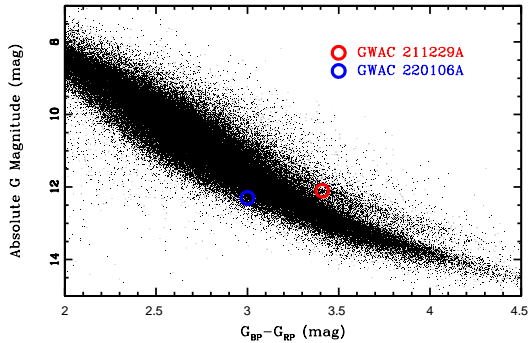}
\caption{CMD of Gaia stars. The host stars at quiescent status are marked by the
red and blue circles for GWAC\,211229A and GWAC\,220106A, respectively.
\label{fig:general}}
\end{figure}

Standard routines in the IRAF package, including bias
and flat-field corrections, were adopted to reduce the raw images taken by the GWAC-F60A telescope.
The light curves were then built by standard aperture photometry and
calibration that is again based on the SDSS catalog through the Lupton
(2005) transformations. After combining the GWAC and GWAC-F60A measurements, 
the final $R-$band light curves are shown in Figure 2 for the two flares. Note
that the effect of reddening can be safely ignored in both cases
throughout the current study, because the extinctions in the Galactic
plane along the line of sight are as low as $E(B-V)=0.08$ mag
and 0.09 mag for GWAC\,211229A and GWAC\,220106A, respectively, 
based on the updated dust reddening map provided by Schlafly \& Finkbeiner (2011).
In addition, based on the hydrogen density around Sun of $n_{\mathrm{H}}=10^6{\mathrm{cm^{-3}}}$ and 
constant dust-to-gas ratio (Bohlin et al. 1978), the relation $E(B-V)\approx053\times(d/\mathrm{kpc})$
results in a rough estimation of $E(B-V)=0.03$ mag and 0.08 mag for GWAC\,211229A and GWAC\,220106A, 
respectively.    
Due to the GWAC’s high cadence of 15 s, the detected peak magnitude is simply
adopted as the real peak brightness of the flare.

\begin{figure}[ht!]
\plotone{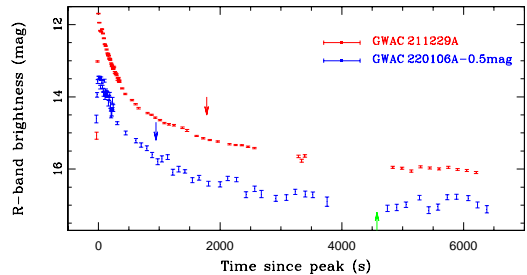}
\caption{$R$-band light curves of GWAC\,211229A (red symbols) and GWAC\,220106A 
(blue symbols) observed by the GWAC cameras and GWAC-F60A telescope. The peak times
correspond to MJD=59577.992299 day and 59585.972041 day for
GWAC\,211229A and GWAC\,220106A, respectively. For both flares, the red and blue downward arrows
mark the start of our time-resolved spectroscopic monitors. The green upward arrow 
marks the beginning of the possible ``chromospheric evaporation'' identified by the blueshifted narrow 
H$\alpha$ emission line in GWAC\,220106A (see Section 5.2 for the details). In both light curve, 
the discontinuity of sampling is due to task scheduling carried out automatically by RAVS (Xu et al. 2020). 
\label{fig:general}}
\end{figure}

\subsection{Spectroscopy and Data Reduction}

Time-resolved long-slit spectroscopy was performed by the NAOC 2.16m telescope (Fan et al. 2016) as soon as possible in the Target of Opportunity mode,
after the discovery and identification of the two flares. 
A log of the spectroscopic observations is presented in Table 2, where $\Delta t$ is the time delay
between the start of the first exposure of spectroscopy and the
trigger time. In total, we have 24 and 13 spectra for GWAC\,211229A and GWAC\,220106A, respectively. 
The epochs of the start of the spectroscopy are marked by the downward arrows in Figure 2.
For each of the flares, a corresponding quiescent spectrum was obtained with the identical instrumental 
setup in the next night.

\begin{table}[h!]
\renewcommand{\thetable}{\arabic{table}}
\centering
\tiny
\caption{Log of Spectroscopic Observations Carried Out by the NAOC 2.16m Telescope. }
\label{tab:decimal}
\begin{tabular}{cccc}
\tablewidth{0pt}
\hline
\hline
ID & Sp. Number & Exposure time (s)  & S/N of H$\alpha$ \\
(1)  &   (2) & (3) & (4)\\
\hline
GWAC\,211229A & \multicolumn{3}{c}{$\Delta t=$30.15min}\\
\cline{2-4}
                            & $1-24$ &  300  & 37.1 (24.6, 78.0)\\ 
                            & quiescent & 900 & \dotfill\\
\hline
GWAC\,220106A & \multicolumn{3}{c}{$\Delta t=$16.30min}\\
\cline{2-4}
                            & $1-5$ &  300  & 41.9 (33.2, 51.4)\\
                            & $6-13$ &  600  & 44.1 (39.0, 47.2)\\
                            & quiescent & $3\times600$ & \dotfill\\
                            \hline
\end{tabular}
\tablecomments{Column (1): the ID of the confirmed transient triggered by the GWAC system. Column (2):  the number series of spectrum. 
Column (3): the exposure time in unit of second. Column (4): the mean value of the 
measured signal-to-noise ratio of the total H$\alpha$ emission line. The minimum and 
maximum values are shown in the bracket (see section 5.1 for the details).}
\end{table}

All spectra were obtained by the Beijing Faint Object Spectrograph and Camera (BFOSC) 
that is equipped with a back-illuminated E2V55-30 AIMO CCD. 
Because we focus on the H$\alpha$ emission line in the current study,
the G8 grism with a wavelength coverage of 5800 to 8200\AA\ was used in the
observations. With a slit width of 1.8\arcsec\ oriented in the south–north direction, the
spectral resolution is 3.5\AA\ as measured from the sky lines,
which corresponds to a velocity of $160\ \mathrm{km\ s^{-1}}$ for the H$\alpha$
emission line. The wavelength calibrations were carried out
with iron–argon comparison lamps.
Flux calibration of all spectra was carried out with observations of Kitt Peak National Observatory standard stars (Massey et al. 1988).
The airmass ranged from 1.1 to 1.4 for GWAC\,211229A and from 1.1 to 1.6 for GWAC\,220106A during the observations.

For each transient,
one-dimensional (1D) spectra were extracted from the raw images by using the IRAF 
package and standard procedures, including bias subtraction and flat-field correction. 
In order to build differential spectra (see Section 3.2 for details),
apertures of both the object and sky emission were fixed in  
the spectral extraction of both object and corresponding standard.
The extracted 1D spectra were then calibrated in wavelength and in
flux by the corresponding comparison lamp and standard stars. 
Using the first spectrum with minimum airmass as a reference, the zero-point of 
wavelength calibration was corrected for each spectrum by an alignment of the sky [\ion{O}{1}]$\lambda$6300 
emission line. With these procedure, the wavelength calibration is $\sim0.1$\AA\ for both flares, which corresponds to a velocity of $\sim5\ \mathrm{km\ s^{-1}}$ at H$\alpha$.
Guaranteed by the fixed object and sky extraction apertures, the differential spectra 
are created by directly subtracting the corresponding quiescent one, and displayed in the left panels in
Figure 3. 

\begin{figure*}[ht!]
\plotone{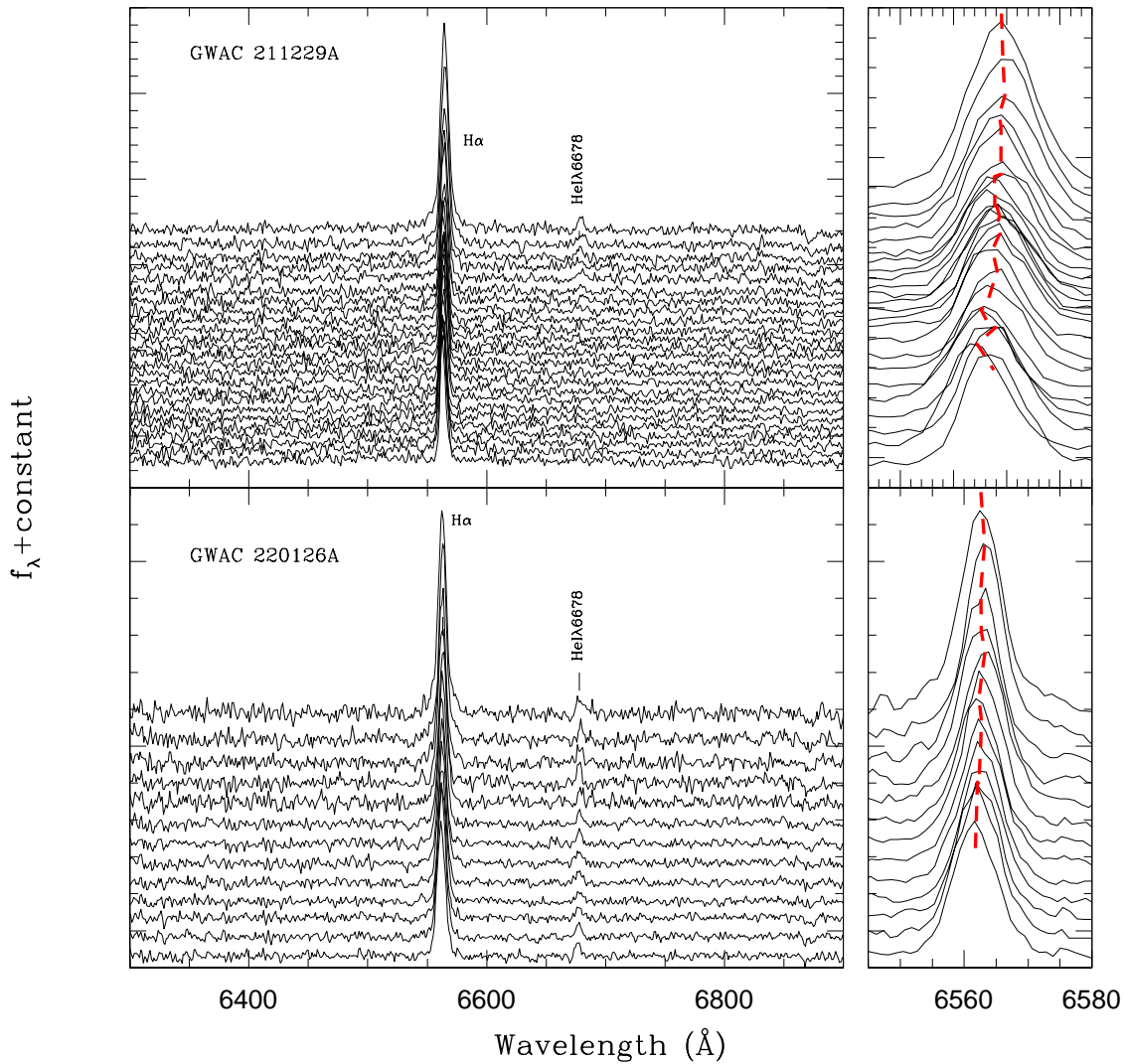}
\caption{\it Left column: \rm Time-resolved differential spectra of GWAC\,211229A and GWAC\,220106A are displayed in the upper and lower panels, respectively. In each 
of the panels, the spectra are sorted with time from top to bottom, and shifted vertically by an arbitrary amount to aid the presentation. The H$\alpha$ and \ion{He}{1}$ \lambda$6678 emission lines are marked on each panel. \it Right column: \rm The same as the left one, but for H$\alpha$ emission line profiles only. The heavy red lines mark the
evolution of the line center of narrow H$\alpha$ component resulted from our spectral 
profile modeling. }
\end{figure*}

\section{Light-curve and Spectral Analyses}

\subsection{Light-curve Analysis}

We model the light curves to estimate the total energy released in the flares by following the 
method adopted in Wang et al. (2021, see also in Xin et al. (2021) and Davenport et al. (2014)). 
The peak relative flux normalized to the quiescent level in the $R-$band, $F_{\mathrm{amp}}$, is at first 
calculated to be 81 and 74 for GWAC\,211229A and GWAC\,220106A, respectively. 
With the calculated $F_{\mathrm{amp}}$,
the lightcurve modelings are shown in Figure 4. In both cases, we fit the rising phase by a linear function: 
\begin{equation}
  \frac{F_{\mathrm{rise}}}{F_{\mathrm{amp}}}=a_0+k_0t
\end{equation}
where $F_{\mathrm{rise}}$ is the relative flux normalized to the quiescent level. The early decaying phase can be
well fitted by a template composed of the sum of a set of exponential components:
\begin{equation}
 \frac{F_{\mathrm{decay}}}{F_{\mathrm{amp}}}=\sum_{i=1}^{N}a_ie^{-\frac{t}{\tau_i}}
\end{equation}
The best fitting returns $N=3$ and 2 for GWAC\,211229A and GWAC\,220106A, respectively. In addition, 
a slow linear decaying is required to account for the tails (i.e., after about 3000 seconds) of both lightcurves.
This linear decaying might be caused by our not long enough monitor and 
result in an overestimation of both equivalent duration (ED) and flaring energy. However, a 
underestimation is certainly returned if the linear decaying is ignored. The modeled values of ED,
with and without the linear decaying phase, are tabulated in Table 1. 

\begin{figure}[ht!]
\plotone{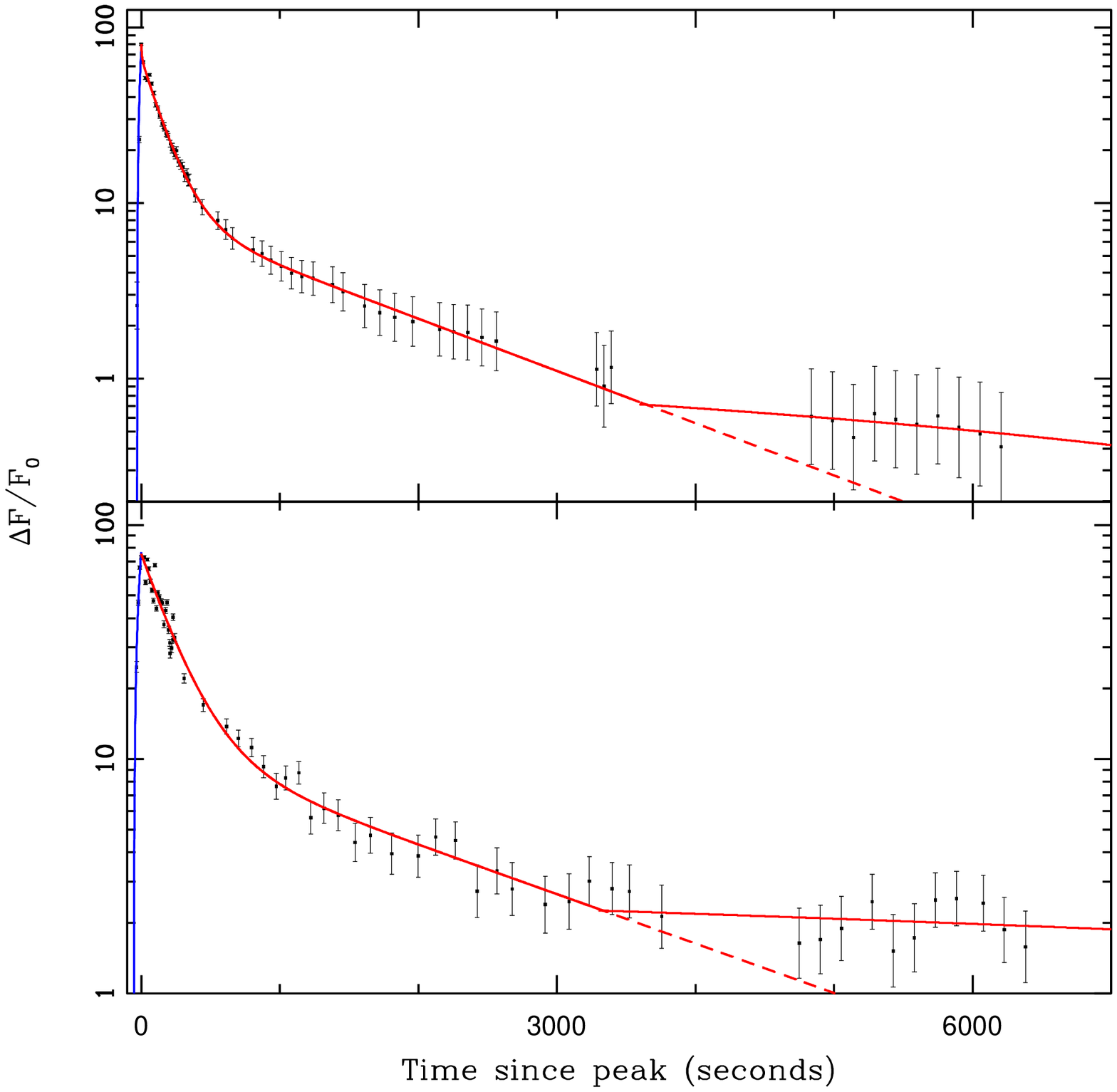}
\caption{Modeling of the light curves of the relative flare flux of GWAC\,211229A (upper
panel) and GWAC\,220106A (lower panel). In each panel, the best fitting
models in the rising phase and decaying phase are denoted by the blue and red lines, respectively.
In each panel, 
the red dashed line denotes a sum of a set of exponential functions which returns a good fit for the early 
decaying phase and a underestimation of the flaring energy. The red solid line denotes the model with an 
additional slow linear decaying at the end of the light curve, which might yield an overestimation of the 
flaring energy.
\label{fig:general}}
\end{figure}

\subsection{Spectral Analysis}

With the differential spectra, we model the H$\alpha$ line profile by a sum of a linear continuum and 
a set of Gaussian function by using the SPECFIT task (Kriss 1994) in the IRAF package. The modelings are 
detailed as follows.  
\begin{itemize}
\item \bf{GWAC\,211229A.} \rm In addition to a narrow Gaussian component, a broad Gaussian component
 is required to properly reproduce the differential H$\alpha$ line profiles in only the first 
 four spectra, which are illustrated in Figure 5. The line width 
 of the broad component is of $\mathrm{FWHM\sim700-800\ km\ s^{-1}}$. A correction of
 $\sigma^2=\sigma^2_{\mathrm{obs}}-\sigma^2_{\mathrm{inst}}$ is applied to the measured line widths, in which an instrumental velocity dispersion of $\sigma_{\mathrm{inst}}=140 \mathrm{km\ s^{-1}}$ is 
 adopted in the correction.
 
\begin{figure}[ht!]
\plotone{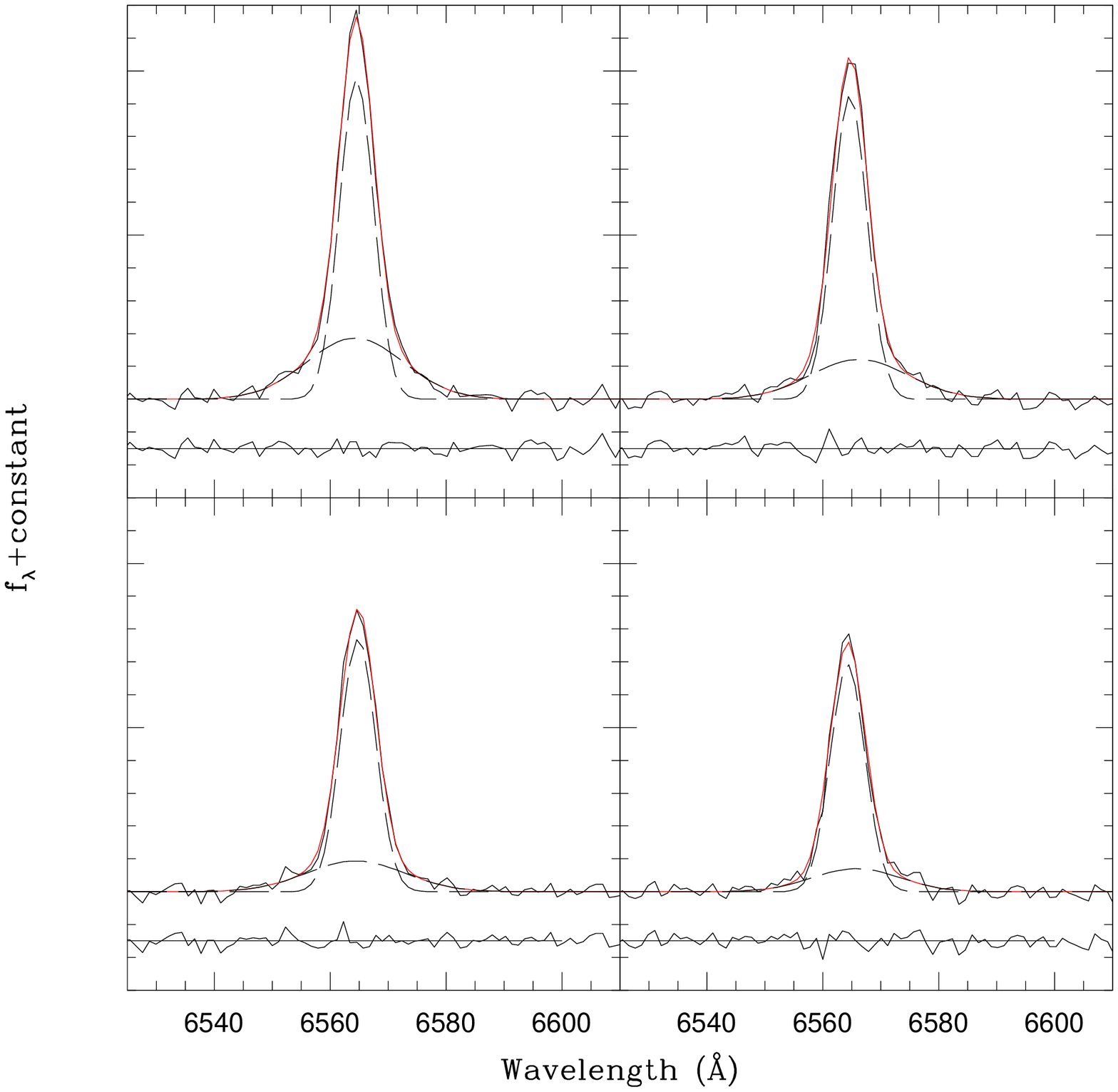}
\caption{An illustration of the line profile modeling using a linear combination of a set of Gaussian functions for the H$\alpha$ emission lines of GWAC\,211229A. The figure only 
shows the first four spectra in which a broad H$\alpha$ component is required to reproduce the observed profiles. In each panel, the modeled local continuum has already been removed from the original observed spectrum. The observed and modeled line profiles are plotted by black and
red solid lines, respectively. Each Gaussian function is shown by a dashed line. The sub-panel underneath each line spectrum presents the residuals between the
observed and modeled profiles.
\label{fig:general}}
\end{figure}

\item \bf{GWAC\,220106A.} \rm Except the last spectrum, the differential H$\alpha$ line 
profiles can be modeled by two Gaussian functions, one is narrow and the other is broad, although the 
necessity of the broad component is questionable in the \#4, \#5 and \#10 spectra.   
The line profile modelings are presented in Figure 6. With the correction of instrumental resolution again, the line width of the broad component is measured to be $\mathrm{FWHM\sim600-700\ km\ s^{-1}}$. 

\begin{figure}[ht!]
\plotone{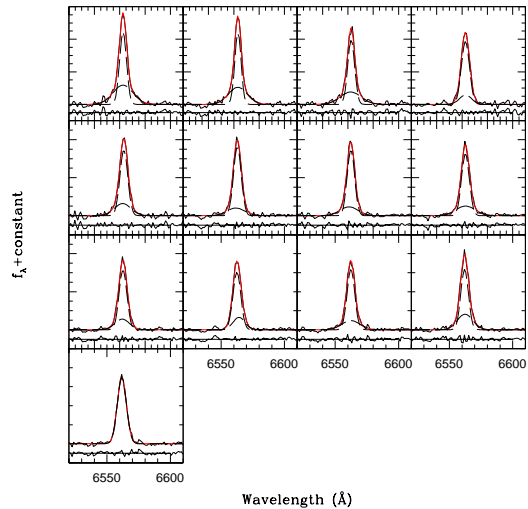}
\caption{The same as in Figure 5, but for GWAC\,220106A. All the 13 spectra are displayed 
in the figure. 
\label{fig:general}}
\end{figure}
\end{itemize}

The results of our line profile modeling are summarized in Table 3. 
The contribution from the corresponding quiescent state is not included in  
the reported fluxes of the H$\alpha$ narrow component. The flux of the quiescent H$\alpha$ emission
is measured to be $f_0=2.54\times10^{-14}$ and $f_0=7.77\times10^{-16}\ \mathrm{erg\ s^{-1}\ cm^{-2}}$ for 
GWAC\,211229A and GWAC\,220106A, respectively.
Column (4) is the line width of the broad H$\alpha$ emission corrected by the instrumental resolution. 
The line shifts are listed in columns (5) and (6) for the narrow and broad components, respectively.
The shifts are calculated from $\Delta\upsilon=c\Delta\lambda/\lambda_0$, 
where $\lambda_0$ and $\Delta\lambda$ are the rest-frame wavelength
in vacuum of a given emission line and the measured wavelength shift of
the line center, respectively.  
Column (7) lists the maximum
projected velocity $V_{\mathrm{max}}$ that is measured from the observed spectrum
at the position where the H$\alpha$ high-velocity blue wing merges
with the continuum. The uncertainties reported in the table corresponds to 
1$\sigma$ significance level due to the modeling.

\begin{table}[h!]
\renewcommand{\thetable}{\arabic{table}}
\centering 
\tiny
\caption{Results of Spectral Measurements and Analysis.}
\label{tab:decimal}
\begin{tabular}{ccccccc}
\tablewidth{0pt}
\hline
\hline
ID & $f\mathrm{(H\alpha_n)}$ & $f\mathrm{(H\alpha_b)}$ &  $\mathrm{FWHM(H\alpha_b)}$  & $\Delta\upsilon(\mathrm{H\alpha_n})$ & $\Delta\upsilon(\mathrm{H\alpha_b})$ & $V_{\mathrm{max}}$\\
 &  \multicolumn{2}{c}{$(\mathrm{10^{-15}erg\ s^{-1}\ cm^{-2}})$} & \multicolumn{4}{c}{($\mathrm{km\ s^{-1}})$} \\
(1)  &   (2) & (3) & (4) & (5) & (6) & (7) \\
\hline
\multicolumn{7}{c}{GWAC\,211229A}\\
\hline
1 &  $72.3\pm4.3$  & $39.0\pm3.7$  & $840\pm90$ &  $78\pm3$  &  $67\pm22$ & 800 \\
2 &  $68.8\pm3.6$  & $25.5\pm3.4$  & $840\pm120$&  $84\pm4$  &  $158\pm36$ & 740 \\
3 &  $58.6\pm2.9$  & $19.8\pm3.1$  & $850\pm110$&  $91\pm4$  &  $70\pm39$  & 683 \\
4 &  $51.7\pm3.2$  & $13.3\pm3.8$  & $740\pm140$&  $67\pm6$  &  $133\pm70$ & 536 \\
5 &  $56.0\pm1.3$  & \dotfill & \dotfill & $75\pm4$  & \dotfill & \dotfill\\
6 &  $49.0\pm1.3$  & \dotfill & \dotfill & $76\pm4$  & \dotfill & \dotfill\\
7 &  $44.5\pm1.2$  & \dotfill & \dotfill & $79\pm4$  & \dotfill & \dotfill\\
8 &  $44.8\pm1.0$  & \dotfill & \dotfill & $47\pm3$  & \dotfill & \dotfill\\
9 &  $42.8\pm1.2$  & \dotfill & \dotfill & $49\pm4$  & \dotfill & \dotfill\\
10&  $40.0\pm0.9$  & \dotfill & \dotfill & $49\pm3$  & \dotfill & \dotfill\\
11&  $40.5\pm1.3$  & \dotfill & \dotfill & $71\pm5$  & \dotfill & \dotfill\\
12&  $45.3\pm1.2$  & \dotfill & \dotfill & $55\pm4$  & \dotfill & \dotfill\\
13&  $41.7\pm1.1$  & \dotfill & \dotfill & $69\pm4$  & \dotfill & \dotfill\\
14&  $41.9\pm1.3$  & \dotfill & \dotfill & $62\pm5$  & \dotfill & \dotfill\\
15&  $39.2\pm1.1$  & \dotfill & \dotfill & $72\pm4$  & \dotfill & \dotfill\\
16&  $40.7\pm1.1$  & \dotfill & \dotfill & $38\pm4$  & \dotfill & \dotfill\\
17&  $37.4\pm1.2$  & \dotfill & \dotfill & $60\pm4$ & \dotfill & \dotfill\\
18&  $41.3\pm1.1$  & \dotfill & \dotfill & $40\pm4$  & \dotfill & \dotfill\\
29&  $38.0\pm1.3$  & \dotfill & \dotfill & $13\pm5$  & \dotfill & \dotfill\\
20&  $43.1\pm1.0$  & \dotfill & \dotfill & $-10\pm4$ & \dotfill & \dotfill\\
21&  $43.7\pm1.2$  & \dotfill & \dotfill & $31\pm4$  & \dotfill & \dotfill\\
22&  $47.5\pm1.1$  & \dotfill & \dotfill & $54\pm3$  & \dotfill & \dotfill\\
23&  $46.1\pm1.2$  & \dotfill & \dotfill & $-28\pm4$ & \dotfill & \dotfill\\
24&  $50.0\pm1.2$  & \dotfill & \dotfill & $45\pm3$  & \dotfill & \dotfill\\                                                                                 
\hline
\multicolumn{7}{c}{GWAC\,220106A}\\
\hline
1 & $30.6\pm2.4$ & $23.4\pm2.3$ & $780\pm70$ & $-9\pm5$ & $-9\pm18$ & 750\\
2 & $27.4\pm1.7$ & $19.5\pm1.7$ & $730\pm50$ & $18\pm4$ & $18\pm18$ & 710\\
3 & $27.0\pm2.2$ & $15.6\pm2.1$ & $820\pm100$& $-9\pm5$ & $-27\pm27$ & 720\\
4 & $30.9\pm7.6$ & $6.9\pm8.1$  & $390\pm100$& $-5\pm10$& $0\pm73$ & 327\\
5 & $26.9\pm3.3$ & $1.1\pm3.3$  & $620\pm100$& $23\pm7$ & $-18\pm41$ & 533\\
6 & $27.7\pm4.5$ & $7.1\pm4.1$  & $680\pm270$& $0\pm5$  & $-55\pm55$ & 625\\
7 & $25.4\pm2.1$ & $8.2\pm2.0$  & $680\pm110$& $-23\pm5$& $-87\pm82$ & 621\\
8 & $25.2\pm2.8$ & $9.6\pm2.6$  & $740\pm140$& $-5\pm5$ & $0\pm32$ & 713\\
9 & $24.3\pm6.8$ & $7.9\pm6.8$  & $520\pm160$& $-9\pm4$ & $-36\pm23$ & 506\\
10& $23.4\pm6.9$ & $7.6\pm6.9$  & $380\pm40$ & $-18\pm6$& $59\pm55$ & 377\\
11& $24.2\pm1.9$ & $8.5\pm1.8$  & $690\pm90$ & $-18\pm4$& $-23\pm27$ & 634\\
12& $23.8\pm12.1$& $12.4\pm11.1$& $570\pm270$& $-32\pm6$& $-18\pm18$ & 577\\
13& $31.3\pm0.5$ & \dotfill   & \dotfill  & $-50\pm3$ & \dotfill & \dotfill \\
\hline
\hline                           
\end{tabular}
\tablecomments{Column (1): the number of spectrum in time series. Columns (2) and (3):  the modeled 
flux in unit of $\mathrm{10^{-15}erg\ s^{-1}\ cm^{-2}}$ of the H$\alpha$ narrow and broad component, respectively.
Each component is denoted by a Gaussian function.
Columns (4): the line width (full width at half maximum) in unit of $\mathrm{km\ s^{-1}}$ of the H$\alpha$  broad component.  
Columns (5) and (6): the bulk velocity shift in unit of $\mathrm{km\ s^{-1}}$ with respect to the rest-frame wavelength of H$\alpha$ line. 
Column (7): the maximum velocity $V_{\mathrm{max}}$ in unit of $\mathrm{km\ s^{-1}}$ of the broad H$\alpha$ line blue wing (see the main text for the details). } 
\end{table}

\section{Results and Discussions}

Figures 5 and 6 show the temporal evolution of the H$\alpha$ emission lines
for GWAC\,211229A and GWAC\,220106A, respectively. At first glance, one can see from the figures that 
both broad and narrow H$\alpha$ emission decay with time in both flares, although 
the narrow H$\alpha$ emission of GWAC\,211229A re-brightened significantly at about 110 minutes after 
the trigger. 
\rm

\begin{figure}[ht!]
\plotone{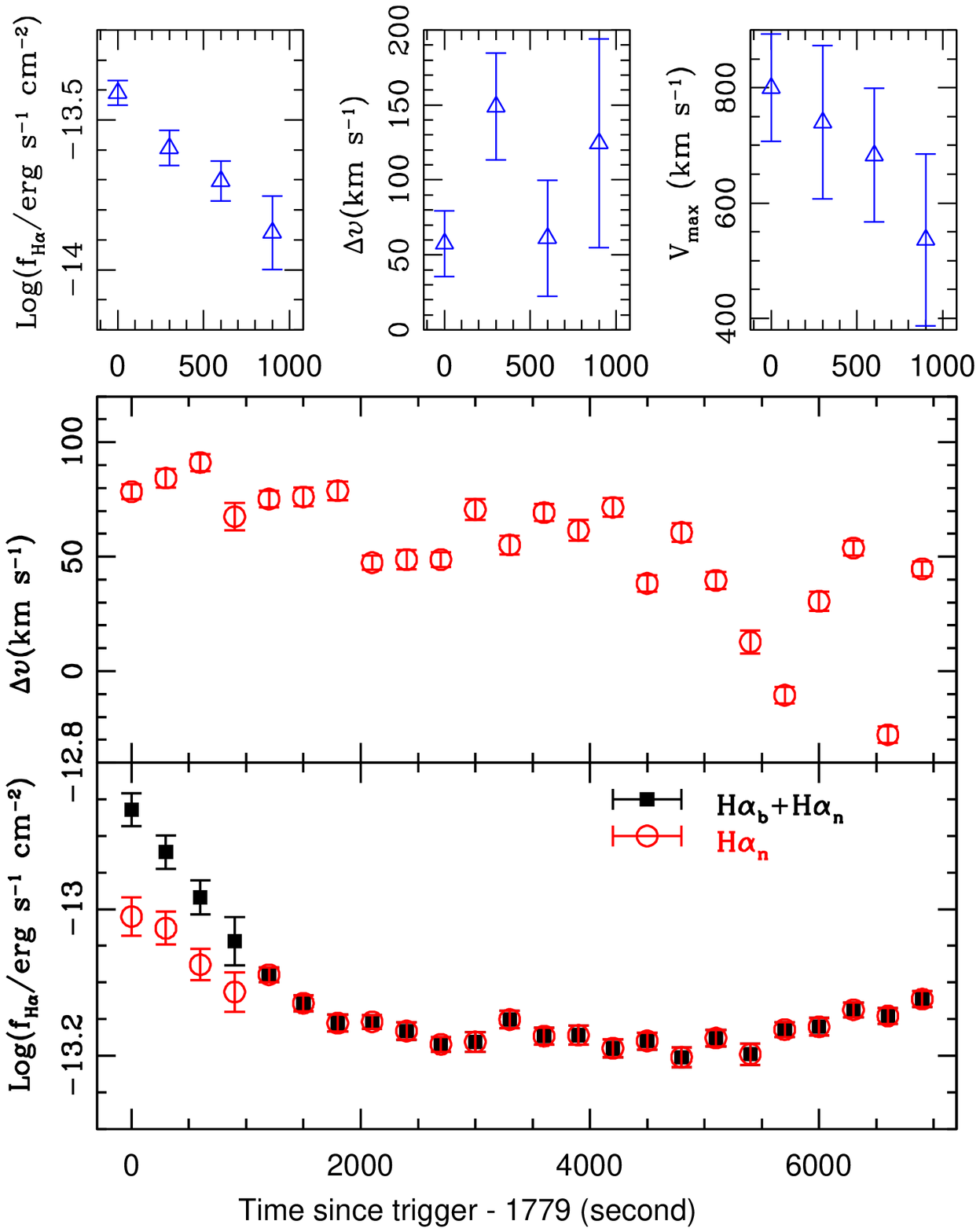}
\caption{\it Top row: \rm Temporal evolution of the broad H$\alpha$ components in GWAC\,211229A. The evolution of line flux, bulk velocity and maximum projected velocity 
at the blue 
line wing are presented from left to right panels. \it Middle and bottom panels: 
\rm The evolution of bulk velocity and line flux of 
the narrow H$\alpha$ emission lines are presented in the middle and bottom panels, respectively. The quiescent H$\alpha$ line flux is included in the figure. 
\label{fig:general}}
\end{figure}

\begin{figure}[ht!]
\plotone{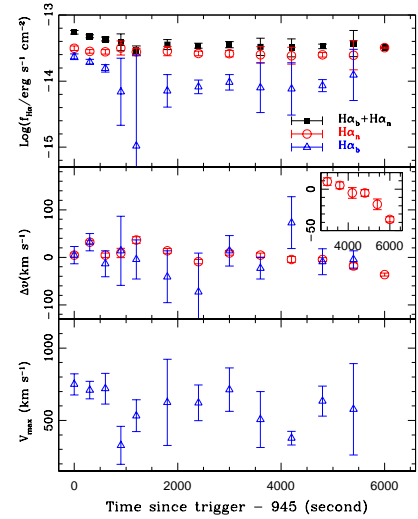}
\caption{Similar as Figure 7, but for GWAC\,220106A. The insert panel shows a gradual increase of the bulk blueshift velocity of the broad H$\alpha$ emission starting at 
about 1.3 hour after the onset of the flare.     
\label{fig:general}}
\end{figure}

\subsection{Physical Origin of the Broad H$\alpha$ Emission}
In both flares, the values of $V_{\mathrm{max}}$ are, in fact, 
obviously larger than the stellar surface escape velocity. 
With $\upsilon_{\mathrm{esp}}=630(M_\star/M_\odot)^{1/2}(R_\star/R_\odot)^{-1/2}\ \mathrm{km\ s^{-1}}$,
the masses and radii tabulated in Table 1 return a $\upsilon_{\mathrm{esp}}=590$ and 
$=580\ \mathrm{km\ s^{-1}}$ for GWAC\,211229A and GWAC\,220106A, respectively.  
This large $V_{\mathrm{max}}$ is unlikely understood in the 
context of either chromospheric evaporation (e.g., Canfield et al. 1990; Gunn et al. 1994; Berdyugina et al. 1999) or condensation (e.g., Antolin et al.
2012; Lacatus et al. 2012; Fuhrmeister et al. 2018; Vida et al.
2019) scenario, although both effects can 
result in an asymmetry of the Balmer emission lines, which has been confirmed in myriad solar observations in 
soft X-ray and EUV. 
The solar observations indicate a velocity of about tens of $\mathrm{km\ s^{-1}}$ in chromospheric evaporation
(e.g., Li et al. 2019), and a velocity no more than $\sim100\ \mathrm{km\ s^{-1}}$ 
(e.g., Ichimoto \& Kurokawa 1984; Asai et al. 2012) in chromospheric condensation. 


Although broad Balmer line emission has been revealed in a batch of 
M dwarfs (e.g., Houdebine et al. 1990; Vida et al. 2016, 2019; Crespo-Chacon et al. 2006;
Koller et al. 2021; Mukehi et al. 2020; Wu et al. 2022; Namekata et al. 2020), its physical origin is 
still under debate. Two possible explanations of line broadening include: (1) Stark (pressure) 
and/or opacity effects and (2) a flaring-associated CME or filament eruption.

\subsubsection{Stark Effects}

Based on analytic approximation and modern radiative-hydrodynamic simulations of the atmospheric response to 
injection of high energy nonthermal and thermal electron beams, the Stark and/or opacity effects have been often proposed to explain 
the symmetrical broad Balmer emission detected at the 
early phase of a flare (e.g., Worden et al. 1984; Hawley \& Pettersen 1991; Johns-Krullet al. 1997; Allred et al. 2006; Paulson et al. 2006; Giziset al. 2013;
Kowalski et al. 2015, 2017; Namekata et al. 2020; Wu et al. 2022).

In a comprehensive study on active M dwarf AD Leonis, Namekata et al. (2020)
proposed a connection between line broadening and non-thermal heating based on
their RADYN numerical simulation and on
the observational fact that the evolution of H$\alpha$ line width is similar to that of the associated
white-light flare during the impulsive and rapid decay phase. 
In spite of a lack of rapid spectroscopy, we reveal long-during  (i.e., 1000-5000 seconds) broad H$\alpha$ emission, extending to the shallow decay phase, in both flares studied here.
This is different from AD Leonis in which the broadening is almost stopped at the end of 
the rapid decay phase.
In addition, in AD Leonis, the line width of H$\alpha$ emission at 1/8 
line peak intensity is found to decrease from 14 to 8\AA\ by the end of the rapid decay phase 
(Namekata et al. 2020). However, a much larger value of $\approx 18$\AA\ is 
obtained in both flares observed by us at the end of the rapid decay phase
(i.e., the beginning of our spectral monitors).

\subsubsection{Flaring-associated CME}

Although a bulk blueshifted emission is usually adopted as a more conclusive indicator of a
CME, the projection effects and the random direction of an filament eruption implies that 
both blueshift and redshift signatures may be observed (see figure 4 in Moschou et al. 2019).
In the two studied flares, a prominence eruption/CME is required to occur on the stellar limb 
to produce the almost bulk zero and slightly redshifted velocities.
Although the foot-points of the two flares are likely be on the visible side of the stellar 
disk due to their strong white-light emission, a limb filament eruption/CME is not 
impossible taking into 
account of the wide distribution of the angle between solar flares and the corresponding CMEs.
Aarnio et al. (2011) and Yashiro et al. (2008), in fact, indicates that the distribution peaks 
at $0\pm45$\degr.   

Beside the limb CME scenario, the slightly redshifted broad emission could be produced
by an absorption due to the erupting filament material. In fact, an obvious absorption at
the blue wing of H$\alpha$ has been identified in a fraction of solar CMEs 
(e.g., Den \& Kornienko 1993; Ding et al. 2003). Finally, a failed filament eruption
could not be entirely excluded to explain the redshifted broad H$\alpha$ emission. 
In this scenario, the erupted filament material is confined by an overlying magnetic field, which finally
results in a downward mass falling (e.g., Drake et al. 2016) and a red asymmetry of the emission line.
The numerical simulations carried out by Alvarado-Gomez et al. (2018) show that a large-scale
dipolar magnetic field of 75G is strong enough for suppressing a CME. In addition to a
red asymmetry, a short-lived blue asymmetry is expected at early time after the onset of a flare.
This blue asymmetry has not been observed in GWAC\,211229A, possibly due to our relatively late 
spectral monitors.
As shown in Figure 8, blueshifted broad H$\alpha$ emission can be, however, briefly detected in 
GWAC\,220106A from 2000 to 3000s after the onset of the flare. 

Without a further statement, the limb CME scenario is preferred for the origin of 
the observed broad H$\alpha$ emission in the subsequent study. 
\rm

\subsubsection{Mass of CME}
 
By following Houdebine et al. (1990, see also in Koller et al. 2021 and Wang et al. 2021), the corresponding CME mass $M_{\mathrm{CME}}$ in each flare can be estimated from the total 
hydrogen mass involved in the CME: 
$M_{\mathrm{CME}}\geq N_{\mathrm{tot}}Vm_{\mathrm{H}}$, where $N_{\mathrm{tot}}$ is the
number density of hydrogen atoms,  $m_{\mathrm{H}}$ the mass of 
the hydrogen atom, and $V$ the total volume. $V$ is related with line luminosity $L_{ji}$ as 
$L_{ji}=N_jA_{ji}h\nu_{ji}VP_{\mathrm{esc}}$, where $N_j$ is the number
density of hydrogen atoms at excited level $j$, $A_{ji}$ the Einstein coefficient
for a spontaneous decay from level $j$ to $i$, and $P_{\mathrm{esc}}$ is the escape probability.
CME mass can therefore be estimated as 
\begin{equation}
  M_{\mathrm{CME}}\geq\frac{4\pi d^2f_{\mathrm{line}}m_{\mathrm{H}}}{A_{ji}h\nu_{ji}P_{\mathrm{esc}}}\frac{N_{\mathrm{tot}}}{N_j}
\end{equation}
where $d$ is the distance and $f_{\mathrm{line}}$ the corresponding line flux. With the limb CME  
scenario, the total flux of the broad H$\alpha$ emission is adopted in the estimation.  
\rm
Due to a lack of $N_3/N_{\mathrm{tot}}$ values in the literature, we estimate $M_{\mathrm{CME}}$ from 
H$\gamma$ line flux that is transformed from the observed H$\alpha$ line flux by
assuming a Balmer decrement of three (Butler et al. 1988). 
Adopting $A_{52}=2.53\times10^6$ (Wiese \& Fuhr 2009)
and $N_{\mathrm{tot}}/N_5=2\times10^9$ estimated from nonlocal thermal
equilibrium modeling (Houdebine \& Doyle 1994a, 1994b) yields a $M_{\mathrm{CME}}\sim4\times10^{18}$ and 
$2\times10^{19}$ g for GWAC\,211229A and GWAC\,220106A, respectively, in which 
a typical value of 0.5 is used for $P_{\mathrm{esc}}$ (Leitzinger et al. 2014).
Both estimated CME masses are in fact in line with the
previously expected range of $10^{14-19}$ g of a stellar CME (Moschou et al. 2019).

We estimate the minimum detectable CME mass $M_{\mathrm{CME,min}}$ for the two flares according to Equation (9) in Odert et al. (2020)
\begin{equation}
  M_{\mathrm{CME,min}}\simeq \frac{\pi R^2_\star m_{\mathrm{H}}N_{\mathrm{H}}}{\mathrm{SNR}\times W[1-e^{-\tau}]}
\end{equation}
where $R_\star$ is the radius of a host star, $m_{\mathrm{H}}$ the mass of hydrogen atom and $N_{\mathrm{H}}$
the column density of a prominence. $W$ and $\tau$ are the geometric dilution factor and optical depth of H$\alpha$ emission line, respectively. SNR is the signal-to-noise ratio of the
emission line. With the SNRs\footnote{In the estimate of S/N ratio of an emission line, the statistic error of the line $\sigma_l$ is determined by the 
method given in Perez-Montero \& Diaz (2003): $\sigma_l=\sigma_c\sqrt{N[1+\mathrm{EW}/(N\Delta)]}$, where 
$\sigma_c$ is the standard deviation of continuum in a box near the line, $N$ the number of pixels used to measure the line flux, 
$\mathrm{EW}$ the equivalent width of the line, and $\Delta$ the wavelength dispersion in units of $\mathrm{\AA\ pixel^{-1}}$. } tabulated in Column (4) of Table 2, $M_{\mathrm{CME,min}}$
is inferred to be $9\times10^{15}$ and $7\times10^{15}\ \mathrm{g}$ for GWAC\,211229A and GWAC\,220106A, respectively, when typical values of $W=0.5$, $N_{\mathrm{H}}=10^{20}\ \mathrm{cm^{-3}}$ and $\tau=10$ (Odert et al. 2020) are adopted in the estimation.

\subsection{Complex Gas Dynamics Implied by Narrow H$\alpha$ Emission?}

Figures 6 and 7 show a temporal variation of radial velocity $\Delta\upsilon$ of narrow H$\alpha$ emission lines in both GWAC\,211229A and GWAC\,220106A. 
The variation of $\Delta\upsilon$ of both flares is further illustrated in the 
right panels in Figure 3 by connecting the modeled line centers of the narrow H$\alpha$ 
components by the heavy red lines.


\subsubsection{GWAC\,220106A}
Although $\Delta\upsilon$ in GWAC\,220106A is around zero at the 
beginning and middle of our spectroscopic monitor, it gradually increases to $\sim-50\ \mathrm{km\ s^{-1}}$
at the end of the monitor.

The maximum rotation velocity of the host star 
is $\upsilon=50(R_\star/R_\odot)(P/\mathrm{day})^{-1}=9(P/\mathrm{day})^{-1}\ \mathrm{km\ s^{-1}}$, where 
the measured radius reported in Table 1 is adopted. It is known for a long time that there is a 
wide distribution (from 0.1-10 day) of the stellar rotation period of M dwarfs 
(e.g., Popinchalk et al. 2021). Based on their X-ray luminosity, 
the stellar activity is found to decrease with the period (e.g., Pizzolato et al. 2003).
In fact, most of the well studied late-type flaring stars in the solar neighborhood
have a rotation period of a couple of days, which corresponds a typical $\upsilon\sin i$ less than
$10\ \mathrm{km\ s^{-1}}$ (e.g., Marcy \& Chen 1992; Pettersen 1980). 
Moreover, a larger H$\alpha$ line shift velocity could be resulted from an absorption due to the 
``slingshot prominence'' that is believed to be supported at or beyond the co-rotation radius by stellar magnetic field.
slingshot prominence is, in fact, identified in many rapidly rotating stars 
(e.g., Collier Cameron \& Robinson 1989; Skelly et al. 2010; Leitzinger et al. 2016). 
Due to its larger radius, slingshot prominence could lead to a line shift velocity comparable 
to the gradual increase of $\Delta\upsilon$ at the end of our spectral monitor.

We argue that the chromospheric evaporation scenario is alternatively a possible explanation of the observed 
gradually blueshifted narrow H$\alpha$ emission line starting at about 1.3 hour after the onset of the flare. 
As shown in Figure 1, it is interesting that this epoch coincides with a rebrightening at the end of the 
light curve. 
In the evaporation scenario, some electrons accelerated by the energy released in the magnetic reconnection (e.g., 
Fletcher et al. 2011; Chen et al. 2020; Tan et al. 2020) heat the 
chromospheric plasma to very high temperature rapidly through Coulomb collisions 
(e.g., Fisher et al. 1985; Innes et al. 1997; Liu et al. 2019; Li 2019; Yan et al. 2021), 
which results in an overpressure in the chromosphere.
The overpressure can drive the plasma either upward (i.e., chromospheric evaporation, 
e.g., Fisher et al. 1985; Teriaca et al. 2003; Zhang ,et al. 2006b; Brosius \& Daw 2015; 
Tian \& Chen 2018) or downward 
(i.e., chromospheric condensation, e.g., Kamio et al. 2005; Zhang et al. 2016a; Libbrecht et al. 2019; 
Graham et al. 2020) motion. The upward and downward velocity depends on the heating flux of the electron beam. 
Generally speaking, the typical upward velocity is tens of $\mathrm{km\ s^{-1}}$ in a ``gentle evaporation'' 
with electron beaming flux $\leq10^{10}\ \mathrm{erg\ cm^{-2}\ s^{-1}}$ (Milligan et al. 2006a; Sadykov et al. 2015; Li et al. 2019). However, in an ``explosive 
evaporation'' with electron beaming flux $\geq3\times10^{10}\ \mathrm{erg\ cm^{-2}\ s^{-1}}$ (e.g., 
Milligan et al. 2006b, Brosius \& Inglis 2017; Li et al. 2017), both a fast upflow 
at hundreds of $\mathrm{km\ s^{-1}}$ and a slow downflow at tens of $\mathrm{km\ s^{-1}}$ can be triggered.  
Although the upflow hot plasma typically favors high temperature emission lines (e.g., 
Tian \& Chen 2018; Polito et al. 2016; Young et al. 2015; Graham \& Cauzzi 2015; Battaglia et al. 2015), cool chromospheric \ion{Si}{4}
\ion{C}{2} and \ion{Mg}{2} emission lines with a blueshift velocity no more than 
$20\ \mathrm{km\ s^{-1}}$ have been identified by Li et al. (2019) in two solar flares.   

\subsubsection{GWAC\,211229A}

In GWAC\,211229A, $\Delta\upsilon$ decreases gradually from $\sim+100\ \mathrm{km\ s^{-1}}$ to around zero.
This measured redshift velocity is indeed somewhat 
larger than the value predicted by the chromospheric evaporation model. 
Alternatively, 
binary scenario is likely able to explain the measured temporal velocity variation
in GWAC\,211229A whose location on the H-R diagram (See Figure 1 for the details) 
implies that it is might be either a  
unresolved binary system or a young main sequence star. Fitting the resulted
temporal velocity variation in GWAC\,211229A by a sinusoidal function returns a 
$\upsilon\sin i=118\ \mathrm{km\ s^{-1}}$ and $P=0.8$day, where $i$ is the orbital inclination.  
By taking velocity $\upsilon=\pi a/P$, where $a$ is separation between the two stars, the Kepler law can be written as $P=9.1\times10^6\bigg(\frac{M_1+M_2}{1M_\odot}\bigg)\bigg(\frac{\upsilon}{1\mathrm{km\ s^{-1}}}\bigg)^{-3}\ \mathrm{day}$,
where $M_1$ and $M_2$ are the masses of the two stars. 
By substituting the fitted radial velocity and period, we find an inclination angle of 
$i\approx47\degr$ by adopting $M_1\approx M_2=0.19M_\odot$.

The quiescent counterpart (i.e., TIC\,177425178) of GWAC\,211229A is bright and monitored by Transiting Exoplanets Survey Satellite (TESS, Ricker et al.  2014). 
With the archival light curve provided in  Mikulski Archive for Space Telescopes (MAST), 
a period analysis is performed by using the PDM (Stellingwerf 1978) task in the IRAF package. 
The period analysis based on the $\theta-P$ diagram is shown in the upper panel of Figure 9. The two minimum values in the plot 
allow us to obtain two periods of $P_1=0.3511(1)$ day (single peak) and $P_2=0.7055(2)$ day (double peaks).  The phase light curves at the 
two periods are displayed in the other two panels in Figure 9. These period modulation  suggests that the object is similar to a
W Ursae Majoris (U Ma)-type binary that is a kind of short-period ($P<1$ day, e.g., Bilir et al. 2005) 
eclipsing binaries generally composed of two dwarfs with comparable temperature and 
luminosity (e.g., Latkovic et al. 2021 and references therein). The two components in the binary are in contact by filling their Roche lobes, which suggests a common envelope
(e.g., Lucy 1968).
In the W UMa scenario, the orbital period inferred from photometry is $P_{\mathrm{orb}}=0.7055(2)$ day that is comparable to the value ($=0.8$day) 
estimated from our spectral analysis. 

\begin{figure}[ht!]
\plotone{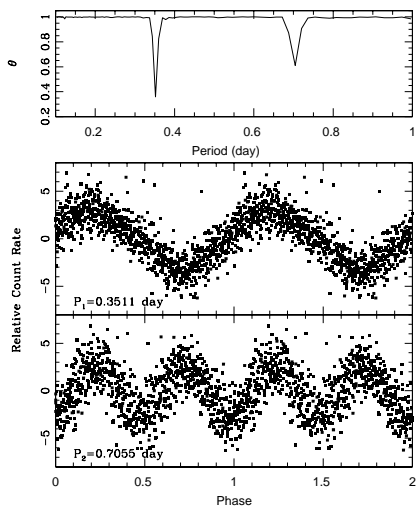}
\caption{Period analysis of TIC\,177425178, the quiescent counterpart of GWAC\,211229A, based on the monitor carried out by TESS. 
\it Upper panel: \rm the $\theta-P$ diagram. There are two minimums corresponding to two periods (single peak and double peaks)
in the light curve. \it Middle and lower panels: \rm the folded phase light curves with a period of 0.3511 (single peak) and 0.7055 (double peaks) day. 
\label{fig:general}}
\end{figure}

\subsection{Flaring Energy Versus CME Mass}

By following Kowalski et al. (2013), the total flaring energy released in $R-$band can be 
estimated from $E_R=4\pi d^2\times f_0\times\mathrm{ED}$, 
where $f_0$ is the quiescent $R-$band flux and 
ED is the equivalent duration. The values of $E_R$ are presented in Table 1 for both flares.
With a bolometric correction of $L_{\mathrm{bol}}/L_R=6$ by assuming a blackbody with 
an effective temperature of $T_{\mathrm{eff}}=10^4$K, the bolometric energy can then be 
estimated as $E_{\mathrm{bol}}\sim(3.1-3.2)\times10^{34}$ and $\sim(7.2-9.6)\times10^{34}$ erg for 
GWAC\,211229A and GWAC\,220106A, respectively. 

We estimate the flaring energy released in X-ray from the X-ray luminosity that can be 
obtained from H$\alpha$ line luminosity according to the relationship (Martinez-Arnaiz et al. 2011; 
Moschou et al. 2019): $\log L_\mathrm{X}=-2.72+1.48\log L_{\mathrm{H\alpha}}-0.93\log R_\star$, 
where $R_\star$ is the stellar radius. The lightcurve modeling finally leads to an estimate of 
$E_\mathrm{X}\approx(3.7-3.8)\times10^{33}$ and $\approx(1.2-1.7)\times10^{33}$ erg for 
GWAC\,211229A and GWAC\,220106A, respectively. 
The ratio of $E_\mathrm{X}/E_{\mathrm{bol}}$ is therefore inferred to be $\sim0.12$ and 
$\sim 0.01-0.02$ for GWAC\,211229A and GWAC\,220106A, respectively. Both values 
are in agreement with the ones $\approx0.01$ revealed by solar observations (e.g.,
Kretzschmar 2011; Emslie et al. 2012).

Figure 10 plots flare bolometric energy $E_{\mathrm{bol}}$ against CME mass $M_{\mathrm{CME}}$
derived in different ways. In addition to the four M-dwarf flares studied in Wang et al. (2021) and in the current study, the figure includes 1) the stellar CME candidates
compiled from 
Moschou et al. (2019, and references therein), in which $M_{\mathrm{CME}}$ is estimated 
through either Doppler shift or X-ray absorption; 2) active G-type giant HR\,9024 studied 
by Argiroffi et al. (2019), who reported a detection of CME by a delayed and
blueshifted \ion{O}{8}$\lambda18.97$\AA\ emission line in time-resolved high-resolution
X-ray spectroscopy; 3) the confirmed solar flare-CME events studied in Yashiro \& Gopalswamy (2009).
A universal bolometric correction of $L_{\mathrm{X}}/L_{\mathrm{bol}}=0.01$ is adopted for 
obtaining $E_{\mathrm{bol}}$ from the energy released in X-ray.  
One can see from the figure that the four M-dwarf flares reinforce the trend that stronger the flaring, higher the CME mass is (e.g., Yashiro et al. 2008; Yashiro \& Gopalswamy 2009; Aarnio et al. 2011; Webb \& Howard 2012; Moschou et al. 2019; Drake et al. 2013). 
Taking into account of the fact that, strictly speaking, the $M_{\mathrm{CME}}$ derived in the four M-dwarf 
flares are lower limits, the four M-dwarf flares tend to lie above the best fit of the solar data. 

The kinetic energy of CME along the line-of-sight (LoS) axis could be naively estimated by $E_{\mathrm{k}}
\geq1/2M_{\mathrm{CME}}\overline{\upsilon}^2$, where $\overline{\upsilon}$ is the mean measured LoS velocity.
With the estimated CME masses and measured velocities, $E_{\mathrm{k}}$ is inferred to be 
$\sim2\times10^{32}$ and $\sim3\times10^{31}$ erg for GWAC\,211229A and GWAC\,220106A, respectively. 
These values are indeed negligible when compared with not only the bolometric radiation energy release, 
and the ones predicted from the solar relationship $\log E_{\mathrm{k}}=(0.81\pm0.85)+(1.05\pm0.03)
\log E_{\mathrm{X}}$ (Drake et al. 2003).   
Taking into account of a drag force done by a strong overlying magnetic field
(e.g., Vrsnak et al. 2004; Zic et al. 2015).
Drake et al. (2016, see also in Alvarado-Gomez et al. (2018)) proposed a CME suppression mechanism that 
is related with a deficient CME kinetic energy.

\begin{figure}[ht!]
\plotone{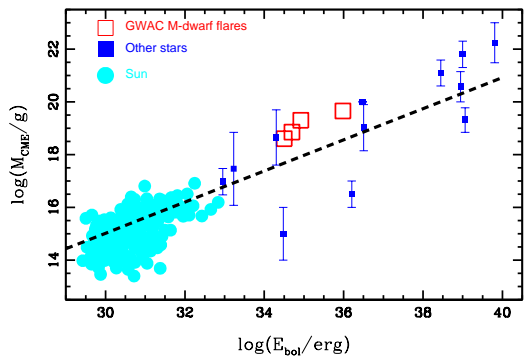}
\caption{CME mass is plotted as a function of flaring bolometric energy. The four 
M-dwarf flares studied by us are denoted by red squares. The stellar CME candidates 
complied in Moschou et al. (2019) and one studied in Argiroffi et al. (2019) are 
shown by the solid blue points. The cyan circles are the solar flare-CME events studied in Yashiro \& 
Gopalswamy (2009). The best fit to the solar events obtained in Drake et al. (2013) is 
presented by a dashed line. 
\label{fig:general}}
\end{figure}

\section{Conclusions}

Two M-dwarf flares are monitored in photometry and time-resolved spectroscopy as soon as possible after the triggers.
Large projected maximum velocity as high as $\sim700-800\ \mathrm{km\ s^{-1}}$ is identified in 
the H$\alpha$ emission in the two flares.
Based on the almost zero bulk velocity, the broadening could be  
ascribed to either Stark effect or a stellar CME occurring at the stellar limb. 
In the latter scenario, the CME mass is estimated to be
$\sim4\times10^{18}$ g and $2\times10^{19}$ g. The temporal evolution of the 
line center of H$\alpha$ narrow emission with a velocity at tens of $\mathrm{km\ s^{-1}}$ could be 
understood by the 
chromospheric evaporation effect in one case and by a binary scenario in the other event. 
By including the four M-dwarf flares studied by us, we show a reinforced trend in which
larger the flaring energy, higher the CME mass is.

\acknowledgments

The authors thank the anonymous referee for his/her careful review and suggestions improving the manuscript significantly. 
The authors are grateful for support from the National Key Research
and Development Project of China (grant 2020YFE0202100). J.W. is supported
by the Natural Science Foundation of Guangxi
(2020GXNSFDA238018) and by the Bagui Young Scholars Program. 
This study is supported by the National Natural Science Foundation of
China (Grants No. 12173009), and by the Strategic
Pioneer Program on Space Science, Chinese Academy of Sciences,
grants XDA15052600 and XDA15016500. This study is supported by the Open Project Program of
the Key Laboratory of Optical Astronomy, NAOC, CAS.
We thank the night assistants
and duty astronomers of the NAOC 2.16 m telescope and the
GWAC system for their instrumental and observational help. Special thanks go to 
the allocated observers for allowing us to interrupt their observations at the 2.16m telescope.
This work has made use of data from the European
Space Agency (ESA) mission Gaia (https://www.cosmos.esa.
int/gaia), processed by the Gaia Data Processing and Analysis
Consortium (DPAC, https://www.cosmos.esa.int/web/gaia/
dpac/consortium). Funding for the DPAC has been provided
by national institutions, in particular the institutions participating 
in the Gaia Multilateral Agreement. This research has made
use of the VizieR catalog access tool, CDS, Strasbourg, France
(doi:10.26093/cds/vizier). The original description of the
VizieR service was published in A\&AS 143, 23. This study uses the 
MAST: Barbara A. Mikulski Archive for Space Telescopes that is a NASA funded project to support and provide 
to the astronomical community a variety of astronomical data archives.
\vspace{5mm}
\facilities{GWAC, GWAC-F60A, NAOC 2.16\,m telescope}
\software{IRAF (Tody 1986, 1992), Python}
%


\end{document}